\documentclass[letterpaper]{JHEP}
\usepackage{epsfig}

\title{Reduction of open membrane moduli}
\author{Eric Bergshoeff\\
    Institute for Theoretical Physics\\
    University of Groningen\\
    Nijenborgh 4, 9747 AG Groningen, The Netherlands\\
    E-mail: \email{bergshoeff@phys.rug.nl}}
\author{Jan Pieter van der Schaar\\
    Michigan Center for Theoretical Physics\\
    Randall Laboratory, Department of Physics, University of Michigan\\
    Ann Arbor, MI 48109-1120, USA\\
    E-mail: \email{jpschaar@umich.edu}}
\abstract{We perform a general reduction of the open
membrane metric in a worldvolume direction of the M5-brane.
Using reduction rules analogous to the bulk, we show that
the open membrane metric leads to the standard
open string metric and open string coupling constant on the
D4-brane only for an ``electric'' reduction in which case the
open membrane metric has no off-diagonal components and the
Born-Infeld curvature tensor is a matrix of rank 2.
Instead, if we perform a general reduction, with nonzero
off-diagonal components of the open membrane metric,
we obtain a rank 4 Born-Infeld tensor corresponding to a
bound state of an open string with an open
D2--brane. Next, we identify and reduce a 3-form
open membrane ``noncommutativity'' tensor on the M5-brane.
This open membrane parameter only reduces to the
open string noncommutativity tensor on the D4-brane
provided we constrain ourselves to an ``electric'' or a ``magnetic''
reduction.}


\preprint{MCTP-01-51 \\ UG-36/01}

\begin{document}
\bibliographystyle{JHEP}

\section{Introduction}

One of the recent
intriguing developments in M-theory has been the insight that
a certain decoupling limit of the elusive M5-brane leads to a (non-commutative)
open membrane theory \cite{Gopakumar-OM,Bergshoeff-OM}.
This open membrane theory, shortly called OM-theory, is
a generalization of the non-commutative open strings found on D-branes
in string theory \cite{Gopakumar-NCOS}.
The geometry of OM-theory is rather unclear at
the moment. In the same way that the non-commutative open strings
are related to a non-commutative geometry one expects that the
open membrane is related to a (non-commutative) loop space with
a (non-associative?) geometry \cite{Bergshoeff-NCM5}.
The non-commutative geometry
of the open string is characterized by a non-commutativity parameter
$\theta^{ab}$. In this paper we will discuss the possibility
that one can introduce a 3-form ``non-commutativity''\footnote{The
parentheses indicate that we expect some (non-associative?) generalization
of a non-commutative geometry in the open membrane case.} tensor
$\Theta^{abc}$ for the open membrane that
is related to $\theta^{ab}$ via dimensional reduction. The possibility
of introducing theta parameters for the open membrane has been
discussed independently in \cite{Swedes}.

The main purpose of this paper is to take the expressions for the open
membrane metric and 3-form non-commutativity parameter and show
under which conditions they reduce to the standard expressions for
the open string metric/coupling constant and non-commutativity parameter.
This work is a generalization of an earlier work by one of us \cite{JPvdS}
where, in order to fix the conformal factor of the open membrane metric,
a specific reduction of the open membrane metric was considered
leading to a rank 2 Born-Infeld matrix on the D4-brane.
In this paper we will analyze in which sense the results of \cite{JPvdS}
can be generalized to the rank 4 case.

Our starting point is the M5-brane.
For some basic M5-brane preliminaries we refer to \cite{JPvdS}.
We consider the following six-dimensional symmetric tensor
defined on a single (Abelian) M5-brane
($\hat{a},\hat{b} \in (0,1,\ldots,5)$)\footnote{We use a mostly plus signature convention
for the metric. The 3-form fields are dimensionless. We use hats
to distinguish the $D=6$ fields and indices from the $D=5$ fields and
indices.}
\begin{equation}
\hat{C}^{\hat{a}\hat{b}} = {1 \over K} \, \left[ (1+{1 \over 12}
\hat{{\cal H}}^2) \, \hat{g}^{\hat{a}\hat{b}}-\, {1\over 4}
(\hat{{\cal H}}^2)^{\hat{a}\hat{b}} \right] \, .
\label{boillat}
\end{equation}
The gauge invariant 3-form field strength
\begin{equation}
\hat{\cal H} \equiv d\hat{b} + \hat{C}
\end{equation}
is defined in terms of a 2-form
gauge field $\hat{b}$ living on the M5-brane and the 3-form
gauge field $\hat{C}$ of $D=11$ supergravity.  The 3-form $\hat{\cal H}$
satisfies the following
nonlinear self-duality equation on the M5-brane \cite{Howe1}:
\begin{equation}
\left. \hat{C}_{\hat{a}} \right.^{\hat{d}} \hat{{\cal H}}_{\hat{d}
\hat{b}\hat{c}} = {\sqrt{-{\rm det}\hat{g}} \over 3!} \epsilon_{\hat{a}
\hat{b}\hat{c}\hat{d}\hat{e}\hat{f}}\hat{{\cal H}}^{\hat{d}\hat{e}\hat{f}} \, .
\label{nlsdeq}
\end{equation}
We have also defined $(\hat{{\cal H}}^2)^{\hat{a}\hat{b}}\equiv
\hat{\cal H}^{\hat{a}\hat{c}\hat{d}}
\left.\hat{\cal H}^{\hat{b}}\right._{\hat{c}\hat{d}}$
and introduced a function $K$ given by
\begin{equation}
K=\sqrt{1+{1\over 24} \hat{{\cal H}}^2} \, .
\label{k}
\end{equation}
In order that the tensor (\ref{boillat}) has the
correct (OM-metric) signature we will only consider the
positive branch of the square root. This implies that  $K\geq 1$.
The symmetric tensor (\ref{boillat}) reduces to the so-called
Boillat metric of nonlinear DBI electrodynamics \cite{Gibbons-West}. We
will refer to this tensor as the M5-brane Boillat metric.

From now on we will assume that the 6-dimensional M5-brane worldvolume is
flat, i.e.~$g^{\hat{a}\hat{b}} = \eta^{\hat{a}\hat{b}}$.
The nonlinear self-duality equation (\ref{nlsdeq}) leads to the following
constraint on $(\hat{\cal H}^4)^{\hat{a}\hat{b}}$:
\begin{equation}
(\hat{\cal H}^4)^{\hat{a}\hat{b}} = {2\over3} \hat{\cal H}^2 \left[
\hat{\eta}^{\hat{a}\hat{b}} + {1\over2} (\hat{\cal H}^2)^{\hat{a}\hat{b}}
\right] \, .
\label{H4constraint}
\end{equation}
Tracing this equation we find
\begin{equation}
{1\over4} \hat{\cal H}^4 = \hat{\cal H}^2 (1+{1\over12}\hat{\cal H}^2)\, .
\label{trH4}
\end{equation}
Using (\ref{H4constraint}) one can verify that the
inverse of $\hat{C}^{\hat{a}\hat{b}}$ is given by\footnote{Because
indices are raised and lowered with the worldvolume metric $\hat{g}_{\hat{a}
\hat{b}}$, we must clearly distinguish between inverse
 metrics and co-metrics, e.g.~$(\hat{C}^{-1})_{\hat{a}\hat{b}} \neq
\hat{C}_{\hat{a}\hat{b}}\equiv \hat{g}_{\hat a\hat c}\hat{g}_{\hat b
\hat d}\hat{C}^{\hat c\hat d}$.}

\begin{equation}
(\hat{C}^{-1})_{\hat{a}\hat{b}} = {1 \over K} \, \left[
\hat{\eta}_{\hat{a}\hat{b}} + \, {1\over 4}
(\hat{{\cal H}}^2)_{\hat{a}\hat{b}} \right] \, .
\label{inverseb}
\end{equation}
We note that the traces of $\hat{C}$ and its inverse are both
equal to $6K$ and that we have, according to
\cite{Gibbons-West}, the remarkable identity ${\rm det}\, (\hat{C}^{-1}
)_{\hat a\hat b} =
{\rm det}\, \hat{g}_{\hat a\hat b}$.
The open membrane co-metric is conformal to the
M5-brane Boillat metric:
\begin{equation}
\hat{G}^{\hat{a}\hat{b}}_{\rm OM} = z \, \hat{C}^{\hat{a}\hat{b}} \, .
\label{defOM}
\end{equation}
The conformal factor $z$ was determined to be equal to
\cite{JPvdS}
\begin{equation}
z = \left( K - \sqrt{K^2 -1} \right)^{-{1 \over 3}} \, .
\label{z(K)}
\end{equation}
Because $K\geq 1$ we find that $0 < z^{-3} \leq 1$
The expression (\ref{z(K)}) for the conformal factor was recently confirmed
using the requirement of deformation independence \cite{Swedes}.

\section{General reduction of the open membrane metric}

To reduce the open membrane co-metric we split the
$D=6$ indices into $\hat{a}=(a,y)$
where $y$ corresponds to a compact direction in the worldvolume of the
M5-brane. Next, we identify the $D=5$ (dimensionless) 2-form and
3-form fields as follows
\begin{eqnarray}
\hat{\cal H}_{aby} &\equiv& {\cal F}_{ab}\, , \nonumber \\
\hat{\cal H}_{abc} &\equiv& {\cal H}_{abc}\, .
\label{redfields}
\end{eqnarray}

As a consequence of the nonlinear self-duality equation in $D=6$
the 3-form ${\cal H}$ and the 2-form ${\cal F}$
are related through a set of nonlinear duality equations given by
\begin{eqnarray}
\left. C_a \right.^d {\cal H}_{dbc} + \left. C_a \right.^y {\cal F}_{bc}
&=& {1\over2} \epsilon_{abcde} {\cal F}^{de}\, , \label{5dnld1} \\
\left. C_y \right. ^d {\cal H}_{dab} + \left. C_y \right.^y {\cal F}_{ab}
&=& -{1\over 3!} \epsilon_{abdef} {\cal H}^{def}\, , \label{5dnld2} \\
\left.C_a\right.^d {\cal F}_{db} &=& -{1\over 3!} \epsilon_{abdef}
{\cal H}^{def} \label{5dnld3} \, .
\end{eqnarray}
The different components of $\hat{C}$ are given by\footnote{
Our conventions slightly differ from those of [6]
in the sense that we use a matrix multiplication
convention after reduction to the D4-brane. This means that
$({\cal F}^2)_{ab}$ in these notes is $-({\cal F}^2)_{ab}$ in [6]
and we will use ${\rm tr}\, {\cal F}^2$ to denote $-{\cal F}^2$ in
[6].}

\begin{eqnarray}
C^{ab} &\equiv& \hat{C}^{ab} = {1 \over K} \, \left[ (1+{1 \over 12}
{\cal H}^2 - {1\over4} {\rm tr} \, {\cal F}^2) \, \eta^{ab}-\, {1\over 4}
({\cal H}^2)^{ab} +{1\over2} ({\cal F}^2)^{ab} \right]\, ,
\label{Cab} \\
C^{ay} &=& {-1\over 4K} \, {\cal V}^a\, , \label{Cay} \\
C^{yy} &=& {1\over K} \left( 1+ {1\over 12} {\cal H}^2 \right) \, \eta^{yy}
\label{Cyy} \, .
\end{eqnarray}
In deriving the expressions (\ref{Cab}), (\ref{Cay}) and (\ref{Cyy}) we have
used that $\hat{\cal H}^2 = {\cal H}^2 - 3 {\rm tr}\,
{\cal F}^2$ and $(\hat{\cal H}^2)^{ab} = ({\cal H}^2)^{ab} - 2
({\cal F}^2)^{ab}$.
Furthermore, we have defined a 5-dimensional
vector ${\cal V}^a$ as follows:
\begin{equation}
{\cal V}^a \equiv \left.{\cal H}^a\right._{cd} {\cal F}^{cd} =
(\hat{\cal H}^2)^{ay} \, . \label{vdef}
\end{equation}
Because of the duality equations this vector is a null eigenvector of
${\cal F}$, i.e.~${\cal V}^a {\cal F}_{ab} = 0$. This implies that
in 5 dimensions, for a generic rank 4 background ${\cal F}$,
the vector ${\cal V}^a$
has only one nonzero component. At first sight it seems that
for a rank 2 background ${\cal F}$ the fact that ${\cal V}^a$
is a null eigenvector of ${\cal F}$ is less restrictive. However, using
the definition (\ref{vdef}) and expressing ${\cal V}^a$ in terms of
${\cal F}$ only, it is not hard to see that ${\cal V}^a$
vanishes in that case. Therefore, a rank 2 reduction implies that
the off-diagonal
vector component
of the open membrane metric vanishes and vice versa \cite{JPvdS}.

One can obtain another set of relations, consistent with
the 5-dimensional duality equations, by reducing
the constraint (\ref{H4constraint}) on $(\hat{\cal H}^4)^{\hat{a}\hat{b}}$:
\begin{eqnarray}
\left( \left.({\cal H}^2)^a\right._c - 2 \left.({\cal F}^2)^a\right._c \right)
\left( ({\cal H}^2)^{cb} - 2 ({\cal F}^2)^{cb} \right) + {\cal V}^a
{\cal V}^b &=& \label{1reducH4}\\
{2\over3}({\cal H}^2 - 3 {\rm tr}{\cal F}^2)
\left( \eta^{ab} - ({\cal F}^2)^{ab} + {1\over 2} ({\cal H}^2)^{ab} \right)
\, ,&&
\nonumber \\
\left[ {1\over3} {\cal H}^2 \eta^{ac} - \left( ({\cal H}^2)^{ac} -
2 ({\cal F}^2)^{ac} \right) \right] {\cal V}_c &=& 0\, ,
\label{2reducH4} \\
{\cal V}^2 - {2\over3}({\cal H}^2 - 3{\rm tr}{\cal F}^2) +
{1\over3} {\cal H}^2 {\rm tr} {\cal F}^2 &=& 0\, .
\label{3reducH4}
\end{eqnarray}

According to \cite{Howe1} the most general solution to the 5-dimensional
duality equations is given by\footnote{Since our main purpose in this paper
is to investigate under which conditions the reduction of the open
membrane leads to open strings, we have solved ${\cal H}$ in terms
of ${\cal F}$. To analyze when the open membrane reduces to open D2-branes
it would be convenient at this point to solve ${\cal F}$ in terms of
${\cal H}$.}
\begin{equation}
{1\over 3!} \epsilon_{abcde} {\cal H}^{cde} = {{ {\cal D}_0 {\cal F}_{ab} +
({\cal F}^3)_{ab}} \over {\sqrt{{\cal D}}}} \, .
\label{soldual}
\end{equation}
We have introduced two 5-dimensional scalars,
${\cal D}$ and ${\cal D}_0$,
which are defined by
\begin{eqnarray}
{\cal D} &\equiv& -{\rm det}(\eta_{ab} + {\cal F}_{ab}) = {\cal D}_0 +
{1\over 8}({\rm tr}\, {\cal F}^2)^2 - {1\over 4} {\rm tr}\,
{\cal F}^4 \, , \label{detn+f} \\
{\cal D}_0 &\equiv&  1-{1\over 2}{\rm tr}\, {\cal F}^2 \, . \label{dnot}
\end{eqnarray}
To avoid an imaginary DBI action we need ${\cal D} > 0$ and
${\cal D}_0>0$\footnote{The critical
electric field limit corresponds to ${\cal D}_0 \downarrow 0$.}.
Furthermore, for electric fields we have
${\cal D}_0 < 1$ whereas magnetic fields correspond to ${\cal D}_0 > 1$.
In the following we will make use of the following identity for
antisymmetric matrices in 5 dimensions:
\begin{equation}
({\cal F}^5)_{ab} = ({\cal D}_0 -{\cal D}) \, {\cal F}_{ab} +
(1-{\cal D}_0)\, ({\cal F}^3)_{ab} \, .
\label{F5eq}
\end{equation}

Using all the previous formulae one can calculate the
following useful tensors, enabling us to eliminate ${\cal H}$ in terms
of ${\cal F}$:
\begin{eqnarray}
{1\over 6} {\cal H}^2 \, \eta_{ab} - {1\over 2}({\cal H}^2)_{ab} &=&
{{\left[ ({\cal D}_0)^2 + ({\cal D}_0 - {\cal D}) \right]
({\cal F}^2)_{ab} + ({\cal D}_0 + 1) ({\cal F}^4)_{ab}} \over {\cal D}}\, ,
\label{H2ab} \\
{1\over 6} {\cal H}^2 &=& (1-{\cal D}_0) + {{4 ({\cal D}_0 - {\cal D})
+  (1-{\cal D}_0)^2} \over {\cal D}} \, ,
\label{H2} \\
{\cal V}^2 &=& {{({\cal D}_0 +1})^2 \over {\cal D}} \, 4
({\cal D}_0 - {\cal D}) \, . \label{V2}
\end{eqnarray}
We note that ${\cal V}^2$ is proportional to ${\cal D}_0 -
{\cal D} = {1\over 4} {\rm tr}\, {\cal F}^4 - {1\over 8}({\rm tr}\,
{\cal F}^2)^2$.
Upon imposing the constraint
${\cal D}_0 - {\cal D}=0$, or equivalently ${\cal V}^2=0$, we find that
${\rm tr}\, {\cal F}^4 = {1\over 2}({\rm tr}\, {\cal F}^2)^2$. By going to
a skew-symmetric basis one can show that this restricts
us to a matrix ${\cal F}$ of rank 2. For this special case we recover the
results of \cite{JPvdS}, in particular ${\cal V}^a = 0$.
For later reference, we
emphasize that the 5-dimensional scalar ${\cal V}^2$ transforms under
6-dimensional Lorentz rotations and therefore the constraint ${\cal D}_0
-{\cal D}=0$ can be removed by performing a 6-dimensional
Lorentz transformation on the M5--brane worldvolume.

The above results can be used to relate the 6-dimensional
Lorentz scalar $K$, see (\ref{k}), to the 5-dimensional Lorentz scalars
${\cal D}$ (\ref{detn+f}) and ${\cal D}_0$ (\ref{dnot}) as follows:
\begin{equation}
K = {{{1\over 2}({\cal D}_0 + 1)} \over {\sqrt{\cal D}}} \, . \label{KinF}
\end{equation}
Although the right hand side of this equation is only manifestly invariant
under 5-dimensional Lorentz transformations, this equation tells us that
it is actually invariant under 6-dimensional Lorentz transformations.
The conformal factor $z(K)$, see (\ref{z(K)}), which is obviously
a 6-dimensional Lorentz scalar, can be written in terms of the 5-dimensional
scalars ${\cal D}$ and ${\cal D}_0$ as follows
\begin{equation}
z^{-3} = { {1\over 2}({\cal D}_0 + 1) - \sqrt{ {1\over 4}({\cal D}_0 - 1)^2
+ ({\cal D}_0 - {\cal D})} \over \sqrt{\cal D}} \, . \label{zindnot}
\end{equation}
Upon a rank 2 truncation (${\cal D}_0={\cal D}$) we find
\begin{equation}
z^{-3} = { {{1\over 2}({\cal D}_0 + 1) - {1\over 2} |{\cal D}_0 - 1|} \over
\sqrt{\cal D}_0} \, . \label{zindnotr2}
\end{equation}
We see that an electric reduction leads to $z^{-3} = \sqrt {{\cal D}_0}$
whereas a magnetic reduction gives $z^{-3} = 1/\sqrt{{\cal D}_0}$.

For our later purposes we give the expression for the 5-dimensional components
of the Boillat co-metric (\ref{boillat}) in terms of the
2-form ${\cal F}$ only:
\begin{equation}
C^{ab} = {1\over \sqrt{\cal D}}
\left[ {\cal D}\, \eta^{ab} + {\cal D}_0 \, ({\cal F}^2)^{ab} +
({\cal F}^4)^{ab} \right] \, . \label{CabinF}
\end{equation}
One can check that the (5-dimensional) inverse of this matrix
is given by
\begin{equation}
(C^{-1})_{ab} = {1\over \sqrt{\cal D}}
\left[ \eta_{ab} - ({\cal F}^2)_{ab} \right] \, , \label{invCabinF}
\end{equation}
which is indeed the D4-brane Boillat metric \cite{Gibbons-Herdeiro},
confirming the result first reported in \cite{Howe1} and more recently
in \cite{Gibbons-West}.

We are now ready to reduce the open membrane metric (\ref{defOM}),
using an appropriate ansatz, and then write the reduced open membrane metric
in
terms of ${\cal F}$ only.
Before actually doing this it is instructive to make a few observations.
The basic result of \cite{JPvdS} was to show that
the conformal factor of the open membrane metric could be
determined  using an uplifting procedure. The analysis of \cite{JPvdS}
was based upon the assumption that the off-diagonal components
of the open membrane metric vanished (i.e.~${\cal V}^2=0$), restricting
the procedure of \cite{JPvdS} to a rank 2 reduction.

We have noted that the difference between the rank 2 and rank 4  cases is
nothing but a 6-dimensional Lorentz rotation. The conformal factor
(\ref{z(K)}) is a six-dimensional Lorentz scalar and hence is
invariant under such a Lorentz rotation and therefore the same for rank 2
and rank 4 reductions. However, this does not imply we should get the
same results in a general rank 4 reduction as we did in a rank 2 reduction.
One should keep in mind that the open membranes ending on the
M5-brane align themselves along the electric directions of the 3-form
background, e.g.~along the 1 and 2 directions, with ${\cal H}_{012} \ne 0$.
This implies that only an ``electric'' rank 2 reduction, i.e.~a
reduction along the 1 or 2 direction, will lead to
open strings ending on the D4-brane. On the other hand, a ``magnetic'' rank 2
reduction, i.e.~a reduction along the 3,4 or 5 direction, will
give open D2-branes ending on the D4-brane \cite{Swedes}.
From this perspective it seems rather unlikely that we will find the open
string metric and coupling constant in a magnetic rank 2 or a general rank
4 reduction. Instead, in a magnetic rank 2 reduction it will presumably
be more appropriate to rewrite everything in terms of the 3-form ${\cal H}$
instead of the 2-form ${\cal F}$ and define a reduction ansatz involving
an open D2-brane metric and an open D2-brane coupling constant (which would
naturally be associated with the open membrane metric Kaluza-Klein scalar,
see also \cite{Swedes}).
In a general rank 4 reduction, i.e.~an oblique reduction along a linear
combination of the (1,2) and (3,4,5) directions,
we expect to find a bound state of an open string with an open D2-brane.
We do not know much about the description of this system but a priori there
is no reason to expect that it can be described by open string parameters
only. The purpose of the analysis below is to check the above sketched
scenario by performing a general rank 4 reduction of the open membrane
parameters and to investigate under which conditions we can identify,
using a reduction ansatz analogous to a bulk M-IIA reduction,
the Seiberg-Witten open string metric and coupling constant.
In Section 3 we will apply a similar analysis to the open string
non-commutativity tensor.

We now proceed and put forward a general reduction ansatz
for the open membrane metric (or co-metric),
in analogy with the bulk M-IIA ansatz like in \cite{JPvdS},
using the conformal factor (\ref{z(K)}).
More specific, we propose the following reduction ansatz for the
open membrane co-metric
$\hat{G}_{\rm OM}^{\hat{a}\hat{b}} \equiv z \, \hat{C}^{\hat{a}\hat{b}}$:
\begin{eqnarray}
\lambda_{\rm os}^{2/3} (G_{\rm os})^{ab} &=& \hat{G}_{\rm OM}^{ab}   \, ,
\label{cOMredos} \\
\lambda_{\rm os}^{-4/3} &=& {\hat{G}_{\rm OM}^{yy} \over \left( 1 -
\hat{G}_{\rm OM}^{ay}
(\hat{G}^{-1}_{\rm OM})_{ay}  \right)}  \, .
\label{cOMcoupling}
\end{eqnarray}
Note that $\hat{G}_{\rm OM}^{ay} (\hat{G}^{-1}_{\rm OM})_{ay} =
\hat{C}^{ay} (\hat{C}^{-1})_{ay} = -{1\over 16K^2} {\cal V}^2 = {\cal D}_0 -
{\cal D}$.
To ensure that reducing the open membrane metric (instead of the inverse
open membrane co-metric) gives the inverse result we require the following
ansatz for the open membrane metric $(\hat{G}^{-1}_{\rm OM})_{\hat{a}
\hat{b}} \equiv z^{-1} \, (\hat{C}^{-1})_{\hat{a}\hat{b}}$:
\begin{eqnarray}
\lambda_{\rm os}^{-2/3} (G^{-1}_{\rm os})_{ab} &=&
(\hat{G}^{-1}_{\rm OM})_{ab} -
{(\hat{G}^{-1}_{\rm OM})_{ay} (\hat{G}^{-1}_{\rm OM})_{by} \over
(\hat{G}^{-1}_{\rm OM})_{yy}}   \, , \label{OMredos} \\
\lambda_{\rm os}^{4/3} &=& (\hat{G}^{-1}_{\rm OM})_{yy}  \, .
\label{OMcoupling}
\end{eqnarray}

The open string metric and coupling we would like to obtain after reducing
the open membrane metric are equal to\footnote{Throughout this paper we will
suppress any closed string coupling dependence, i.e.~$g_s=1$.}
\cite{Seiberg-Witten}:
\begin{eqnarray}
(G^{-1}_{\rm os})_{ab} &=& \eta_{ab} - ({\cal F}^2)_{ab} \, ,
\label{osmetric} \\
\lambda_{\rm os} &=& \sqrt{\cal D} \, .
\label{SWGos}
\end{eqnarray}
We observe that the open string coupling constant is not
a 6-di\-men\-sio\-nal Lorentz invariant object, i.e.~its value changes when
transforming from a rank 2 to a rank 4 background using a 6-dimensional
Lorentz rotation.

Let us focus our attention on the open string coupling $\lambda_{\rm os}$, as
defined by the open membrane metric (\ref{OMcoupling}). We find
\begin{equation}
\lambda_{\rm os}^{4/3} = (\hat{G}^{-1}_{\rm OM})_{yy} =
z^{-1} (\hat{C}^{-1})_{yy}
= \left( { {1\over 2}({\cal D}_0 + 1) - \sqrt{ {1\over 4}({\cal D}_0 - 1)^2
+({\cal D}_0 -{\cal D})} \over \sqrt{\cal D}} \right)^{1/3} \sqrt{\cal D} \, ,
\label{oscdnot}
\end{equation}
which is obviously not equal to the expression (\ref{SWGos}). In fact, any
attempt to modify the ansatz (\ref{OMcoupling}) (to include for example
arbitrary functions of $\lambda_{\rm os}$), or modify the conformal
factor $z(K) \neq 1$, see eq.~(\ref{z(K)}), will fail to reproduce
(\ref{SWGos}) for the
simple reason that we need to obtain a function of ${\cal D}$ only.
As we already discussed, the conformal factor $z$ has to be a 6-dimensional
Lorentz scalar and therefore it can never be a function of ${\cal D}$ only,
as is confirmed by eq.~(\ref{KinF}).
The only exception is the case $z(K)=1$, when we find
$(\hat{G}^{-1}_{OM})_{yy} = (\hat{C}^{-1})_{yy} = \sqrt{\cal D}$,
which is unsatisfactory because it does
not reproduce the required scaling of the open membrane metric in the
OM-theory decoupling limit \cite{Gopakumar-OM,Bergshoeff-OM} Furthermore,
the
reduction ansatz that would be required does not follow the bulk ansatz.
Any failure to reproduce the open string
coupling also means that the open string metric (\ref{osmetric}) is not
reproduced using the proposed ansatz (\ref{OMredos})\footnote{One does
of course always find something conformal to the open string metric.}.

To remind the reader, when imposing the constraint ${\cal D}_0 - {\cal D}=0$
in (\ref{oscdnot}), we reproduce the open string coupling
$\lambda_{\rm os}=\sqrt{{\cal D}_0}$, as well as the open string metric,
defined through (\ref{OMredos}) \cite{JPvdS}, but only when we consider an
electric reduction\footnote{Taking the square root in (\ref{oscdnot}) we
obtain the {\it absolute} value of ${\cal D}_0 -1$, which is equal to
$1-{\cal D}_0$ in an electric reduction.
In that case we exactly find $\lambda_{os}=\sqrt{{\cal D}_0}$.
On the other hand, for a rank 2 magnetic reduction, with ${\cal D}_0 > 1$,
we find $\lambda_{os}=({\cal D}_0)^{1/4}$.}.
What happens when we
perform a 6-dimensional Lorentz rotation is that the open string coupling as
defined by (\ref{OMcoupling}) will transform into (\ref{oscdnot}), whereas
the rank 4 Seiberg-Witten open string coupling equals (\ref{SWGos}).
Only when we truncate to rank 2 do the two expressions coincide.
Another way to put this is that when $z(K) \neq 1$ the Seiberg-Witten
open string coupling (\ref{SWGos}) can not be related to the
Kaluza-Klein scalar of the open membrane metric because it
does not appropriately transform under 6-dimensional Lorentz transformations.

We conclude that only an electric rank 2 reduction
leads to open strings. Instead, a magnetic rank 2 reduction or
a general rank 4 reduction naturally involves open D2-branes.
One cannot expect that these reductions lead to
open string moduli only. The analysis of this Section shows that,
when open D2-branes are involved, it is not possible to view the open string
coupling constant as the Kaluza-Klein scalar of an open membrane metric.

\section{General reduction of the open membrane theta parameter}

Besides the open string metric and open string coupling one needs the open
string non-commutativity parameter to define the non-commutative open string
theory (NCOS).
The open string non-commutativity
parameter $\theta^{ab}$ has an analogue in the D3-brane DBI theory as the dual
Maxwell field $P^{ab}$. According to \cite{Gibbons-Herdeiro} these
objects are related in the following way
\begin{equation}
\theta^{ab} = {-1 \over \sqrt{\cal D}}
P^{ab} \, , \label{noncomP}
\end{equation}
where the dual Maxwell field $P^{ab}$ equals
\begin{equation}
P^{ab} = {1 \over \sqrt{\cal D}}
\left( {\cal D}_0 {\cal F}^{ab} + ({\cal F}^3)^{ab} \right) \, . \label{dualM}
\end{equation}
This expression for the dual Maxwell field $P^{ab}$ can actually be written as
(using (\ref{CabinF}) and (\ref{F5eq}))
\begin{equation}
P^{ab} =  {\cal F}^{ad} \, C^{b}_{d} \, , \label{PinCF}
\end{equation}
which means that $\theta^{ab} = - {\cal F}^{ad} \, {(G_{\rm os})}^{b}_{d}$.

The question arises whether one can similarly find a 3-form tensor on the
M5-brane that corresponds to an appropriately generalized notion of
non-commutativity. Theta parameters of rank 3 from a
different point of view have also been discussed in e.g.~\cite{Theta3}.
One property of such a 3-form tensor should be that it reduces to
$\theta^{ab} \propto P^{ab}$ on the D4-brane.
Motivated by the expression (\ref{PinCF}) we first introduce the
following 3-form tensor on the M5-brane which is
made out of the 3-form $\hat{\cal H}$ and {\it one} Boillat co-metric:
\begin{equation}
\hat{P}^{\hat{a}\hat{b}\hat{c}} \equiv \hat{\cal H}^{\hat{a}\hat{b}\hat{d}}
\hat{C}^{\hat{c}}_{\hat{d}} = {1 \over K} \, \left[ (1+{1 \over 12}
\hat{{\cal H}}^2) \, \hat{\cal H}^{\hat{a}\hat{b}\hat{c}} -\, {1\over 4}
(\hat{{\cal H}}^3)^{\hat{a}\hat{b}\hat{c}} \right]
 \, . \label{Pnoncom}
\end{equation}
Because of the self-duality condition (\ref{nlsdeq}) we have
$\hat{P}_{\hat{a}\hat{b}\hat{c}} = {1\over 3!} \epsilon_{\hat{a}
\hat{b}\hat{c}\hat{d}\hat{e}\hat{f}}\hat{{\cal H}}^{\hat{d}\hat{e}\hat{f}}$.
Upon reduction, using results of the previous section (most importantly
(\ref{CabinF}) and (\ref{F5eq})), we
find the following 2-form
\begin{equation}
\hat {P}^{yab} \equiv P^{ab} =  {1\over {\sqrt{\cal D}}}
\left[ {\cal D}_0 \, {\cal F}^{ab}
+ ({\cal F}^3)^{ab}\right] \, , \label{Preduc}
\end{equation}
which is indeed equal to (\ref{dualM}). In order that this tensor
exactly reduces to the
non-commutativity parameter $\theta^{ab}$ we need to fix its conformal
factor. Assuming a rank 2 electric reduction, i.e.~${\cal D}=
{\cal D}_0$ and $z^{-3}=\sqrt{{\cal D}_0}$, it is easy to see that this
conformal factor has to be equal to $z^3$ (see (\ref{z(K)})):
\begin{equation}
\theta^{ab} = -z^3 \, P^{ab} \, . \label{thetazP}
\end{equation}

Because we had to introduce the conformal factor $z^3$, see (\ref{zindnot}),
we conclude that a magnetic rank 2 reduction or a general rank 4 reduction
of the three-form $z^3 \, \hat{P}^{\hat{a}\hat{b}\hat{c}}$  yields the
open string non-commutativity tensor (\ref{noncomP}) only up to a conformal
factor. The explanation for this obstruction is the same
as in our discussion of the reduction of the open membrane metric in the
previous Section. These reductions also lead to open D2--branes and therefore
one should presumably not expect to obtain the theta parameter of open
strings only. Note that the 3-form
(\ref{thetazP}) does not even reproduce the open string non-commutativity
parameter in a magnetic rank 2 reduction. In any reduction of an M5-brane
3-form we will get a 2-form and a 3-form on the D4-brane, which should be 
dual to each other. If the M5-brane 3-form is really the generalized open
membrane theta parameter we expect that the 2-form that arises after
a rank 2 electric or magnetic reduction gives the open string 
non-commutativity tensor.

There is another argument to discard the
3-form $-z^3\, \hat{P}$ (\ref{thetazP}) as the open membrane theta
parameter. By definition, the 3-form $\hat{P}$ satisfies a {\it nonlinear}
self-duality condition.
This implies that it does not scale homogeneously in the
OM-theory limit $\ell_p \rightarrow 0$ (which is directly related to
$\hat{\cal H}$ not scaling homogeneously, see \cite{Bergshoeff-OM}). This
does not conform
with our expectation for a generalized M5-brane non-commutativity tensor;
in the OM-theory decoupling limit we expect to find a fixed non-commutativity
scale in all directions on the M5-brane. 
We therefore conclude that the 3-form (\ref{thetazP})
cannot be the generalized open membrane theta parameter we are looking for,
even though it does give the open string non-commutativity tensor in an
electric rank 2 reduction.

Fortunately, there exists another 3-form tensor $\hat{W}$,
first put forward in  \cite{Swedes}, that also reduces to
the open string non-commutativity tensor $\theta^{ab}$ under
an electric rank 2 reduction. The tensor $\hat{W}$ has
the special feature that it reduces to $\theta$ (in a rank 2 reduction)
{\it without} the need to introduce an extra conformal
factor $f(z)$. The absence of such a conformal factor implies that both the
electric and the magnetic rank 2 reduction lead to the
open string non-commutativity tensor on the D4-brane.

The 3-form tensor $\hat{W}$ is made out of the 3-form
$\hat{\cal H}$ and {\it two} Boillat co-metrics \cite{Swedes}:
\begin{equation}
\hat{W}^{\hat{a}\hat{b}\hat{c}} \equiv \hat{\cal H}^{\hat{a}\hat{d}\hat{e}} \,
\hat{C}^{\hat{b}}_{\hat{d}}\,
\hat{C}^{\hat{c}}_{\hat{e}} = \hat{P}^{\hat{a}\hat{b}\hat{d}} \,
\hat{C}^{\hat{c}}_{\hat{d}}\, . \label{OMtheta}
\end{equation}
Expressed in terms of the 3-form $\hat{\cal H}$ only we find the expression
(using a  constraint on $(\hat{\cal H}^5)^{\hat{a}\hat{b}\hat{c}}$ that can
be deduced from (\ref{H4constraint}))
\begin{equation}
\hat{W}^{\hat{a}\hat{b}\hat{c}} = (1+{1\over 6} \hat{\cal H}^2)
\hat{\cal H}^{\hat{a}\hat{b}\hat{c}} - {1\over 2}
(\hat{\cal H}^3)^{\hat{a}\hat{b}\hat{c}} \, . \label{WinH}
\end{equation}
Note that the expressions (\ref{Pnoncom}) and (\ref{WinH})
for the tensors $\hat{P}$ and $\hat{W}$, respectively, in terms of
$\hat{\cal H}$, only differ in the numerical factors. Actually, the
three 3-forms $\hat P$, $\hat W$ and $\hat {\cal H}$ 
are related in the following way:
\begin{equation}
\hat{W}^{\hat{a}\hat{b}\hat{c}} = 2K\, \hat{P}^{\hat{a}\hat{b}\hat{c}} -
\hat{\cal H}^{\hat{a}\hat{b}\hat{c}}  = 2K \, ^{\star}\hat {\cal H}^{\hat a
\hat b\hat c} - \hat {\cal H}^{\hat a\hat b\hat c}
\, . \label{WinPH}
\end{equation}

Under the assumption that the
off-diagonal terms in the Boillat co-metric vanish, i.e.~${\cal D}=
{\cal D}_0$, we find that the 3-form $\hat{W}$ reduces to
\begin{equation}
\hat{W}^{aby} = \hat{\cal H}^{ady} \,
\hat{C}^{b}_{d} \, \hat{C}^{y}_y = \hat{P}^{ab} \, \hat{C}^{y}_y    \, .
\label{Wreduce1}
\end{equation}
This exactly reproduces the open string non-commutativity tensor
$\theta^{ab}$ for the rank 2 case (electric or magnetic)\footnote{Remember
that $\hat{C}^{y}_y={1\over \sqrt{{\cal D}_0}}$.}, without the need to fix the
conformal factor (except for a minus sign). We next wish to check
what happens to this result beyond the rank 2 case. Using the antisymmetry of
$\hat{W}$, we find that
\begin{equation}
\hat{W}^{aby} = \hat{W}^{yab} \equiv \hat{\cal P}^{yad} \,
\, \hat{C}^{b}_d \, .
\label{Wreduc}
\end{equation}
We already know the expressions (\ref{Preduc}) for $\hat{\cal P}^{yad}$
and (\ref{CabinF}) for $\hat{C}^{b}_d$. By repeatedly using the identity
(\ref{F5eq}) for antisymmetric matrices in 5 dimensions we find
\begin{equation}
\hat{W}^{yab} = {1\over {\cal D}} \left[ ({\cal D}_0^2 + ({\cal D}_0 -
{\cal D}) ) {\cal F}^{ab} + ({\cal D}_0 +1) ({\cal F}^3)^{ab} \right] \, ,
\label{WinF}
\end{equation}
which does not reproduce the open string non-commutativity tensor
(\ref{noncomP}). One can check that when imposing ${\cal D}={\cal D}_0$
the expression reproduces (\ref{noncomP})\footnote{One has to use
the fact that the identity (\ref{F5eq}) for antisymmetric matrices in
5 dimensions reduces to a constraint on ${\cal F}^3$ (relating it to
${\cal F}$). This works for both the electric and magnetic rank 2 reduction.
Unlike in the reduction of the open membrane metric there is no square
root involved which could lead to a different result for the electric and
magnetic cases.}, as we already concluded using eq. (\ref{Wreduc}).
Like in the previous Section the explanation for this result is that a
general rank 4 reduction will give a bound state of an open D2-brane and 
an open string ending on the D4-brane. 
The 3-form obtained after a magnetic rank 2 reduction should naturally be
related to a 3-form open D2-brane ``non-commutativity'' tensor on the 
D4-brane \cite{Swedes}.
We conclude that the tensor $\hat{W}$ only reduces to the open string
non-commutativity tensor under a rank 2 electric or magnetic reduction.

Although the property of $\hat{W}$ to reduce to the open string theta
parameter for both electric and magnetic rank 2 is perhaps preferable,
it is not enough to conclude unambiguously that $\hat{W}$ is the open
membrane theta parameter. In order to establish this more convincingly 
we should verify whether the tensor $\hat{W}$ has the expected behavior in the
OM-theory decoupling limit.
In \cite{Swedes} it was shown that the 3-form $\hat{W}$ on the M5-brane is 
unique in the sense that it is the only 3-form that satisfies a {\it linear} 
self-duality condition with respect to the open membrane metric. 
The discussion below will only focus on the behavior of the 3-form $\hat{W}$ 
in the OM-theory limit (see also \cite{Swedes}).
Introducing the correct dimensions $[length]^3$ and rewriting in terms of
the open membrane co-metric (\ref{defOM}) we (re-)define
\begin{equation}
\hat{\Theta}^{\hat{a}\hat{b}\hat{c}} \equiv - \, \ell_p^3 \, \hat{W}^{\hat{a}
\hat{b}\hat{c}} = - \, \ell_p^3 \, z^{-2} \,
\hat{g}^{\hat{a}\hat{k}} \hat{\cal H}_{\hat{k}\hat{l}\hat{m}} \,
(\hat{G}_{OM})^{\hat{b}\hat{l}} \, (\hat{G}_{OM})^{\hat{c}\hat{m}} \, .
\label{dimtheta}
\end{equation}
We want to check that this tensor gives a fixed homogeneous
non-commutativity parameter, relative to the Planck
length $\ell_p$, in the OM-theory limit $\ell_p \rightarrow 0$.
In the OM-theory limit the conformal factor $z^{-2}$ and the open membrane
co-metric scale homogeneously as follows
\begin{equation}
z^{-2} \sim (\ell_p /\ell_g) \quad , \quad (\hat{G}_{OM})^{\hat{a}\hat{b}}
\sim (\ell_g /\ell_p)^2 \, , \label{zGscale}
\end{equation}
where we introduced $\ell_g$ as the fundamental length-scale of OM-theory
\cite{Gopakumar-OM, Bergshoeff-OM}.
Turning on a nontrivial constant 3-form background naturally induces a $3+3$
split and in the OM-theory decoupling limit the (bulk) metric and the 3-form
$\hat{\cal H}$ scale (inhomogeneously) as follows
($\alpha, \beta \in (0,1,2)$ and  $i,j \in (3,4,5)$)
\begin{eqnarray}
\hat{g}^{\alpha\beta} &\sim& 1 \quad , \quad \hat{g}^{ij} \sim
(\ell_g /\ell_p)^3 \, , \nonumber \\
\hat{\cal H}_{012} &\sim& 1 \quad , \quad \hat{\cal H}_{345} \sim
(\ell_p /\ell_g)^3 \, .
\label{OMscaling}
\end{eqnarray}
Using these scalings one can check that $\hat{g}^{ak}
\hat{\cal H}_{\hat{k}\hat{l}\hat{m}}$ is actually fixed ($\sim 1$)
in all directions (homogeneous) in the OM-theory limit. This means we end
up with the following homogeneously fixed expressions of $\hat{\Theta}$, see
(\ref{dimtheta}), in the OM-theory limit $\ell_p \rightarrow 0$
\begin{eqnarray}
\hat{\Theta}^{012} &\sim& \ell_g^3 \, , \label{theta012} \\
\hat{\Theta}^{345} &\sim& \ell_g^3 \,  \, . \label{theta345}
\end{eqnarray}
This is exactly what one would expect of a generalized M5-brane
non-commutativity tensor in the OM-theory limit \cite{Swedes}.

We conclude that there exist two 3-form
tensors on the M5-brane, $- z^3 \,\hat P$ and $\hat \Theta = -\ell_p^3\hat W$,
that both reduce to the open string non-commutativity
tensor under an electric rank 2 reduction. One of them, $\hat \Theta$, 
also reduces to the
open string non-commutativity tensor using a magnetic rank 2 reduction.
Both tensors do not reduce to the open string non-commutativity tensor in a
general rank 4 reduction. An OM-theory scaling argument \cite{Swedes} (and
possibly the fact that only $\hat \Theta$ gives the open string
non-commutativity tensor in an electric {\it and} magnetic reduction)
shows that only the 3-form tensor $\hat \Theta$
can be identified with the open membrane generalized non-commutativity
parameter. The precise expression for this theta parameter
is given by
\begin{equation}
\hat{\Theta}^{\hat{a}\hat{b}\hat{c}} = - \, \ell_p^3 \, z^{-2} \,
\hat{g}^{\hat{a}\hat{k}} \hat{\cal H}_{\hat{k}\hat{l}\hat{m}} \,
(\hat{G}_{OM})^{\hat{b}\hat{l}} \, (\hat{G}_{OM})^{\hat{c}\hat{m}} =
- \, \ell_p^3 \, \left( (1+{1\over 6} \hat{\cal H}^2)
\hat{\cal H}^{\hat{a}\hat{b}\hat{c}} - {1\over 2}
(\hat{\cal H}^3)^{\hat{a}\hat{b}\hat{c}} \right)
\, . \label{ThetainH}
\end{equation}
It is important to realize that we can now study this theta parameter
without considering a particular limit (although generically the M5-brane
theory will not be decoupled from the bulk). For example, in
\cite{Bergshoeff-NCM5} a large $h$ (which is the variable parameterizing
nonlinearly self-dual constant solutions, see \cite{Bergshoeff-NCM5}) 
limit was studied in which a commutator (in the magnetic directions) 
describing a non-commutative loop space
proportional to $h^{-1}$ was found. Going outside the limit one expects
to find corrections to the $h^{-1}$ behavior that make it nonsingular
for $h\rightarrow 0$. We expect to find something like
${h \over 1+h^2 \ell_p^6}$ such that $\hat \Theta \rightarrow 0$
for $h \rightarrow 0$ and $\hat \Theta \sim h^{-1}$ for
$h\rightarrow \infty$. This is what happens in the
non-commutative D-brane cases. These and possibly other interesting
considerations related to (\ref{ThetainH}) will be left for a future
investigation.

\section{Conclusions}

We have performed a general rank 4 reduction of the open membrane metric.
The difference with the rank 2 reduction, discussed by one of us
\cite{JPvdS}, is that the rank 4 reduction contains an extra parameter
which can be associated with an angle that describes the reduction
of the open membrane with respect to a fixed background. The electric rank 2
reduction of \cite{JPvdS} corresponds to a zero angle and a reduction of the
open membrane to an open string. On the other hand, the generic
rank 4 reduction corresponds to a nonzero angle and leads to open D2--branes
as well. We have shown in this work that for those reductions
the open string coupling constant can {\it not} be interpreted as the
Kaluza-Klein scalar of an open membrane metric.

We also discussed the introduction of a
generalized non-commutativity tensor on the M5-brane, see also
\cite{Swedes} for an independent discussion.
We showed that there are two 3-form tensors that reduce to the open string
non-commutativity tensor under an electric rank 2 reduction.
Only one of them, i.e.~the one of \cite{Swedes}, has the
correct behavior in the OM-theory decoupling limit. This tensor
also reduces to the open string non-commutativity tensor under a magnetic rank
2 reduction. It would be interesting to work out this magnetic reduction
in more detail and analyze how the appropriate open D2-brane parameters
are related to the different components of the open membrane metric.

In analogy with open strings, it is natural to conjecture that
the 3-form open membrane ``non-commutativity'' tensor is somehow related to
3-point functions of the open membrane.
It would be interesting to see what the geometric properties are of a
space deformed by such a 3-form tensor and whether it has something to do
with a non-associativity of the M5-brane worldvolume. 
So far, not much progress has
been made in describing such a non-associative geometry although we expect
a relation with non-commutative loop space \cite{Bergshoeff-NCM5}.

Although all open membrane or OM-theory
parameters have now been identified, i.e.~the OM-metric
\cite{Bergshoeff-OM, Swedes, JPvdS} and more recently the OM-theta
parameter \cite{Swedes}, it is unknown what the
microscopic (open membrane) origin of these objects is. This in contrast with
D-branes where the open string 2-point function gives us both the
open string metric and the theta parameter \cite{Seiberg-Witten}.
We hope that the future will see some progress in our
understanding of the microscopic degrees of freedom of OM-theory.
Without doubt this will lead to a better understanding of M-theory.

\acknowledgments
It is a pleasure to thank David Berman, Ulf Gran, Bengt E.W.~Nilsson
and Per Sundell for interesting
discussions. This work was started during a visit of E.B.~to the
Michigan Center for Theoretical Physics. He wishes to thank the
people there for their hospitality.

\newpage

\bibliography{OBMfinal}

\end{document}